\documentclass[pdflatex,sn-mathphys-num]{sn-jnl}

\usepackage{graphicx}%
\usepackage{multirow}%
\usepackage{amsmath,amssymb,amsfonts}%
\usepackage{amsthm}%
\usepackage{mathrsfs}%
\usepackage[title]{appendix}%
\usepackage{xcolor}%
\usepackage{textcomp}%
\usepackage{manyfoot}%
\usepackage{booktabs}%
\usepackage{algorithm}%
\usepackage{algorithmicx}%
\usepackage{algpseudocode}%
\usepackage{listings}%
\usepackage{multirow}
\usepackage{array}
\usepackage{makecell}
\usepackage[table]{xcolor}
\usepackage{graphicx}
\usepackage{subcaption}
\usepackage[version=4]{mhchem}
\usepackage{arydshln}

\theoremstyle{thmstyleone}%
\theoremstyle{thmstyletwo}%

\theoremstyle{thmstylethree}%

\begin{document}

\title[Article Title]{XRDiff: Crystal Structure Prediction from Powder X-Ray Diffraction Data Using Diffusion Models}


\author[1]{\fnm{Nofit} \sur{Segal}}

\author[2]{\fnm{Mingda} \sur{Li}}

\author*[3]{\fnm{Benjamin Kurt} \sur{Miller}}\email{bkmi@meta.com}

\author*[1]{\fnm{Rafael} \sur{Gómez-Bombarelli}} \email{rafagb@mit.edu}

\affil*[1]{\orgdiv{Department of Materials Science and Engineering}, \orgname{MIT}, \orgaddress{\city{Cambridge},  \state{MA}, \country{USA}}}

\affil[2]{\orgdiv{Department of Nuclear Science and Engineering}, \orgname{MIT}, \orgaddress{\city{Cambridge},  \state{MA}, \country{USA}}}

\affil[3]{\orgdiv{FAIR}, \orgname{Meta}, \city{San Francisco}, \state{CA}, \country{USA}}


\abstract{Determining the crystal structure of a material from its powder X-ray diffraction (PXRD) pattern is a central challenge in materials science. PXRD is an accessible and widely used characterization technique, yet recovering the atomic structure from diffraction data requires solving an underdetermined inverse problem due to the loss of phase information. Generative modeling can provide a prior over atomic structure and learn the mapping from PXRD patterns to crystal structures via simulated structure-spectrum pairs. We present XRDiff, a diffusion model that recovers crystal structures from PXRD given either the stoichiometry or, in a more challenging setting, the elemental constituents and total number of atoms in the unit cell. We evaluate on datasets where each stoichiometry has multiple polymorphs and all polymorphs of a given composition are held out together, ensuring that high performance reflects genuine use of the diffraction signal. XRDiff achieves strong structure recovery rates on simulated benchmarks, indicating that the model learns a spectrum-to-structure mapping precise enough to differentiate between polymorphs. To address generalization to experimental data, we compare a full-spectrum encoding against an encoding based on peak descriptors. The peak-based encoding generalizes substantially better, outperforming even a model trained on full spectra with augmentations fitted to the experimental noise distribution. These results demonstrate that representations robust to the noise and artifacts present in real-world PXRD offer a practical and scalable path toward closing the simulation-to-experiment gap, enabling zero-shot crystal structure solution from experimental PXRD with full or partial chemical composition input.}





\maketitle

\section{Introduction}\label{intro}

The crystal structure of a material governs its physical and chemical properties, often making accurate structure determination a prerequisite for understanding and designing functional materials. Powder X-ray diffraction (PXRD) is the most widely used characterization technique for this purpose, owing to its low cost and accessibility \cite{hammond2015basics, bragg1913reflection, dinnebier2015powder}.

However, structure determination from PXRD faces several compounding challenges. The technique compresses three-dimensional structural information into a one-dimensional spectral pattern, causing peak overlap and ambiguity in peak identification \cite{harris1996crystal}. Small distortions to the unit cell can produce discontinuous changes, such as peak splitting \cite{hammond2015basics}, introducing a non-smooth structure-signal relationship \cite{segal2026loss}. Distinct polymorphs may yield strongly correlated patterns \cite{david2008structure, holder2019tutorial, nakai2025crystal}, making differentiation difficult despite its practical importance \cite{omee2025polymorphism}. Current structure solution methods, such as Rietveld Refinement \cite{rietveld1969profile}, require an initial model whose quality strongly influences the outcome. While modern structure solution workflows incorporate automated workflows and database-assisted initialization, the refinement itself remains a local optimization procedure and is therefore extremely sensitive to initialization and data quality \cite{biwer2025spotlight}. Hence, its success can vary considerably with structural complexity and is very limited when no close structural prototypes are available \cite{biwer2025spotlight, harris1996crystal, harris2003structural}. Consequently, current methods remain tools for materials identification rather than the discovery of novel phases.

We focus on the following problem: \textit{given a powder X-ray diffraction pattern and known chemical composition, recover the full 3D crystal structure}. We refer to this as crystal structure prediction (CSP) from PXRD. We also consider a partial composition setting in which the model is given only the constituent atom types and the total number of atoms in the unit cell.

\begin{figure}[t]
    \centering
    \includegraphics[width=\linewidth]{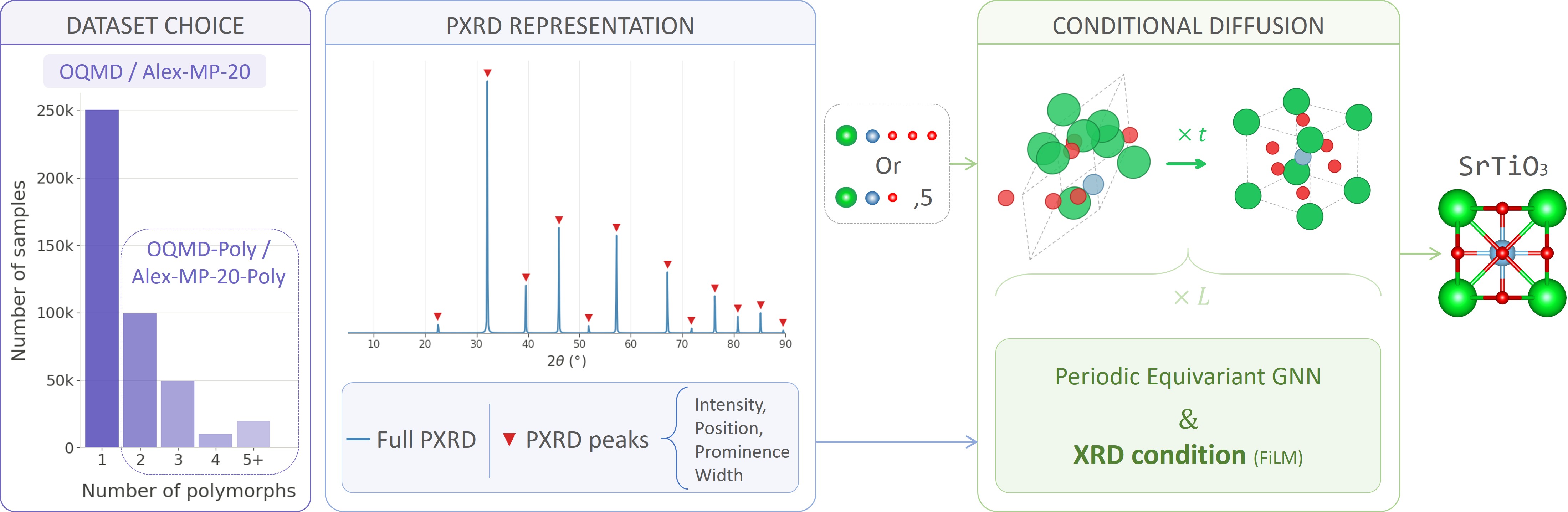}
    
    \caption{Overview of the PXRD-conditioned crystal structure prediction pipeline. \textbf{Left}: Distribution of polymorphs in the OQMD dataset. We screen out unique compositions (OQMD-Poly / Alex-MP-20-Poly) to ensure the model cannot trivially reconstruct structures without leveraging the XRD signal. \textbf{Center}: The PXRD pattern of \ce{SrTiO3} is used as the conditioning signal, represented in two ways: (1) a full PXRD vector of discretized intensities over the $2\theta$ range, and (2) PXRD peak features: intensity, position, prominence, and width of the top-20 peaks by intensity. Both representations perform comparably on clean computational data, but the peak-based representation is significantly more robust to experimental noise. \textbf{Right}: A diffusion model conditioned on the PXRD pattern and composition (either full stoichiometry or atom types plus unit-cell atom count) iteratively denoises a random structure to generate the predicted crystal structure.}
    \label{fig:gen_img}
\end{figure}

Here we present XRDiff, a PXRD-conditioned diffusion model for CSP that recovers the full 3D atomic structure from its 1D powder diffraction pattern given full or partial chemical composition input (Figure~\ref{fig:gen_img}). XRDiff is evaluated under a polymorph-only protocol, where all polymorphs of a given composition are held out together during evaluation, preventing the model from exploiting composition-to-structure shortcuts. This stricter setting ensures that high match rates genuinely reflect the model ability to leverage diffraction information.

A key challenge in learning from PXRD data is the simulation-to-experiment gap: experimental artifacts such as background noise, peak broadening, and instrument-specific distortions hinder generalization from clean simulated training data, which is more abundant and therefore better suited for training. Augmentation-based approaches address this by simulating corruptions during training, but require substantial computational overhead and a representative set of experimental patterns to perform experiment-specific parameter fitting \textemdash yet such a dataset is often unavailable and may not generalize to other experimental conditions. We instead adopt a peak-featurized representation, encoding each diffraction pattern by peak positions, intensities, prominences, and widths. By discarding artifact-contaminated background while retaining the physically meaningful signal, this representation generalizes to arbitrary experimental conditions without any target-dataset fitting, achieving better zero-shot generalization to experimental data than both the full intensity representation and augmentation-based training. We validate this by evaluating XRDiff, trained solely on simulated patterns, against experimental structures from the RRUFF database \cite{lafuente20151}.

Taken together, XRDiff advances CSP from PXRD along three axes: a polymorph-aware evaluation protocol that targets the discovery setting 
(determining the structure of novel compounds not available in databases); a peak-featurized input representation that narrows the simulation-to-experiment gap without augmentation pipelines; and a partial composition setting that moves toward realistic laboratory conditions. XRDiff code is available at \url{https://github.com/learningmatter-mit/xrdiff}.

\section{Background} \label{sec:background}
\textbf{Crystallography} involves determining crystal structures by analyzing diffraction patterns of X-ray, neutron, or electron beams \cite{hammond2015basics, bragg1913reflection}. \textbf{Powder crystallography} is a critical sub-branch that addresses the common practical scenario where a sample consists of many small, randomly oriented crystalline grains rather than a single crystal \cite{dinnebier2015powder, david2008structure}. Since growing large single crystals is often difficult or impossible \cite{harris1996crystal}, powder methods are widely used, though they introduce additional mathematical complexity due to the loss of phase of scattered waves, known as \textbf{the phase problem} \cite{hammond2015basics, hauptman1991phase, egami2012underneath}. 

\textbf{Structure determination} from powder diffraction is an iterative, expert-guided process \cite{giacovazzo2002fundamentals}. It can largely be divided into the determination of lattice parameters and assignment of crystal symmetry and space group through the indexing process, followed by extracting fractional atomic coordinates from Bragg peak intensities to produce a structure model \cite{david2008structure, harris1996crystal, harris2003structural}. Following that, \textbf{Rietveld refinement} is carried out to fit the observed diffraction profile by optimizing the parameterized structural model to minimize the difference between observed and calculated patterns \cite{rietveld1969profile, harris1996crystal, harris2003structural}.

\textbf{Experimental PXRD} patterns contain \textit{artifacts} arising from two sources: the measurement apparatus and the sample itself. Instrument-induced artifacts include counting statistics that introduce intensity noise, as well as systematic peak broadening from beam divergence and detector resolution \cite{hammond2015basics}. Sample-induced artifacts include preferred orientation \textemdash where grains are not truly randomly oriented \textemdash causing systematic intensity distortions; small grain sizes causing broadened peaks (Scherrer broadening); and residual lattice strain shifting peak positions \cite{holder2019tutorial, chandra1999analysis}. Together, these effects further make the mapping from experimental PXRD to structure more ambiguous than in the simulated case, and a model trained purely on clean simulated patterns may fail to generalize.

\section{Related Work}\label{sec:related_work}
Machine and deep learning methods have increasingly been explored to accelerate different stages of crystal structure determination from diffraction data. Early efforts focused on extracting crystallographic information directly from diffraction patterns, including classification of crystal systems and space groups \cite{andrejevic2026alphadiffract, suzuki2020symmetry, corriero2023crystalmela} and regression of lattice or unit cell parameters \cite{andrejevic2026alphadiffract, shu2025machine, gomez2023convolutional}. Beyond direct prediction tasks, several works have investigated how diffraction information can guide or improve structure optimization and refinement strategies \cite{parackal2024identifying, lee2023creation, segal2026loss}. 

Related approaches have also addressed CSP from PXRD by explicitly enumerating candidate Wyckoff configurations consistent with predicted space groups and lattice parameters \cite{parackal2024identifying}. Additionally, retrieval-based approaches have aligned PXRD patterns and crystal structures in a shared latent space for structure identification \cite{xu2026kan}. A complementary line of work predicts charge density from PXRD patterns and stoichiometry \cite{guo2024towards}.

More recently, generative models have emerged for solving CSP from PXRD. These approaches include variational autoencoders that learn latent correspondences between PXRD patterns and crystal structures \cite{riesel2024crystal}, contrastive learning frameworks that align diffraction and structure representations \cite{li2025powder, lai2025end}, language-model-based approaches \cite{choudhary2025diffractgpt}, and diffusion-based generative models that take lattice parameters as input and predict atomic coordinates only \cite{yu2026equivariant}. Concurrently, a flow-matching framework emphasizing robustness to experimental diffraction data through PXRD representation \cite{li2026experimental}. Several of these methods also explore composition-free settings, relaxing the assumption that the chemical composition is known a priori \cite{riesel2024crystal, choudhary2025diffractgpt}. A related work addresses nanoparticle structure determination from PXRD and stoichiometry \cite{guo2025ab}.

\section{Results}\label{results}
We evaluate XRDiff on two data regimes. Section~\ref{sec:comp_results} presents 
results on clean computational data, assessing structure reconstruction quality. 
Section~\ref{sec:exp_results} evaluates generalization to experimental data after 
training exclusively on computational data, examining the simulation-to-experiment 
gap and strategies for bridging it.

\subsection{Computational data} \label{sec:comp_results}
\begin{figure}[t]
    \centering
    \begin{subfigure}[t]{0.98\textwidth}
        \centering
        \includegraphics[width=\linewidth]{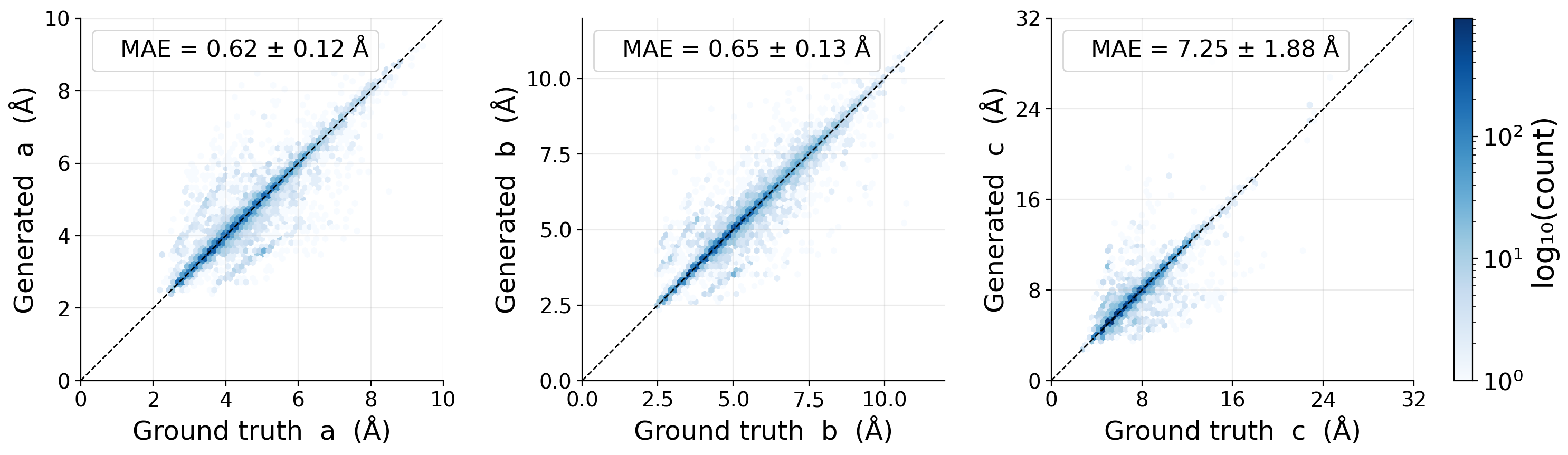}
        \caption{Alex-MP-20-Poly}
        \label{subfig:lat_par_alex_full}
    \end{subfigure}
    \begin{subfigure}[t]{0.98\textwidth}
        \centering
        \includegraphics[width=\linewidth]{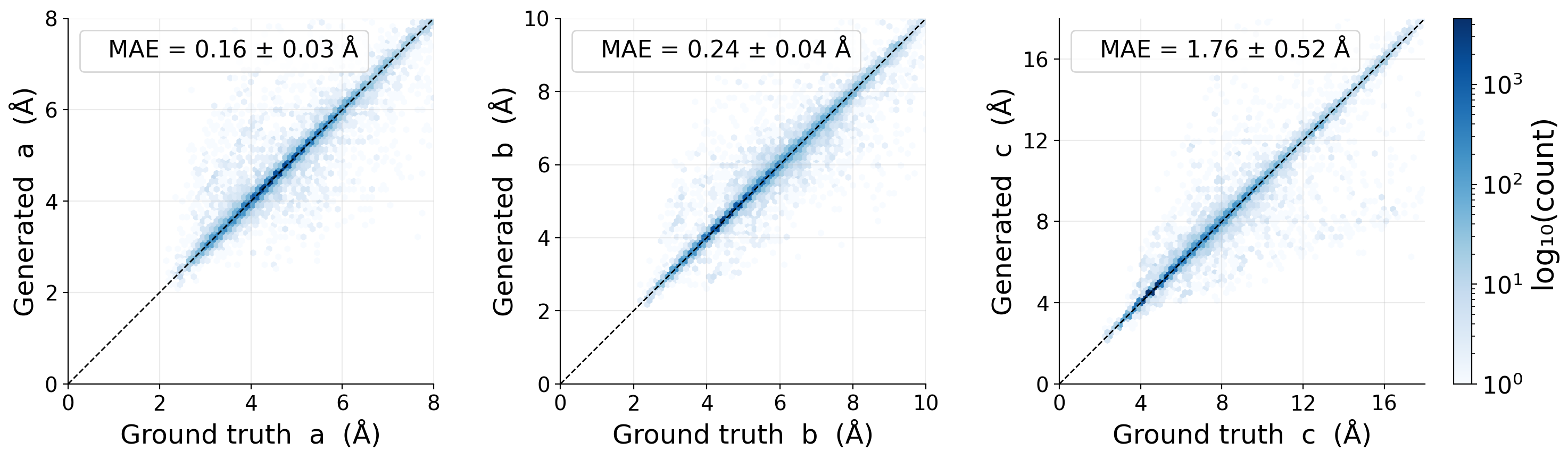}
        \caption{OQMD-Poly}
        \label{subfig:lat_par_oqmd_full}
    \end{subfigure}

    \caption{\textbf{Parity plots for lattice length prediction by XRDiff on the Alex-MP-20-Poly and OQMD-Poly test sets, with point colors denoting local scatter density on a logarithmic scale.} Comparison between ground-truth and generated lattice lengths $a$, $b$, and $c$ for XRDiff conditioned on the full PXRD vector representation and full stoichiometry. The dashed line denotes perfect agreement. The logarithmic color scale allows both dense regions near the parity line and sparse regions with larger prediction deviations to remain visible. Reported values correspond to the mean absolute error (MAE) and the standard error of the mean across the test set.}
    \label{fig:lat_par}
\end{figure}

The match rate performance across datasets and input settings is described in 
Section~\ref{sec:match_rate}, and presented in Figure~\ref{fig:mr} and Tables~\ref{tab:RMSD_comparison},~\ref{appndx:amd}. 
The accuracy of generated structures lattice parameters is described in 
Section~\ref{sec:lattice_eval}, presented through parity plots in Figure~\ref{fig:lat_par}
and MAE Table~\ref{tab:lattice_mae_comparison}
The match rate as a function of structural properties including unit-cell 
size, volume, and crystal system symmetry is analyzed in 
Section~\ref{sec:mr_analysis}, Figure~\ref{fig:fails}. 
A qualitative examination of representative polymorph pairs and 
their corresponding generated structures and diffraction patterns is provided in
Section~\ref{sec:qualitative}, Figure~\ref{fig:polymorphs_Li-Fe-Ni-Mn-O}.

\subsubsection{Structure Reconstruction Match Rate} \label{sec:match_rate}
\begin{figure}[t]
    \centering
    \begin{subfigure}[t]{0.98\textwidth}
        \centering
        \includegraphics[width=\linewidth]{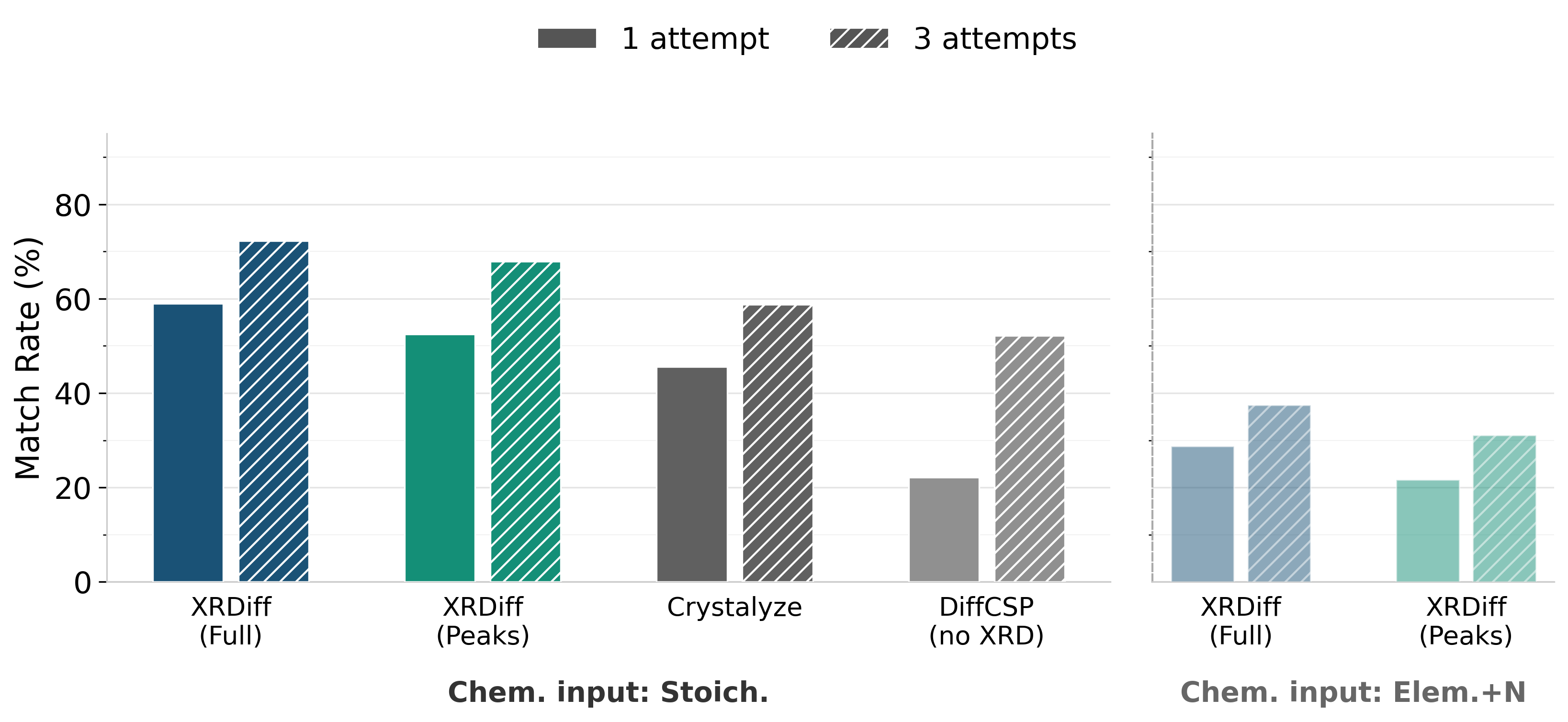}
        \caption{Alex-MP-20-Poly}
        \label{subfig:mr_alex}
    \end{subfigure}
    \begin{subfigure}[t]{0.98\textwidth}
        \centering
        \includegraphics[width=\linewidth]{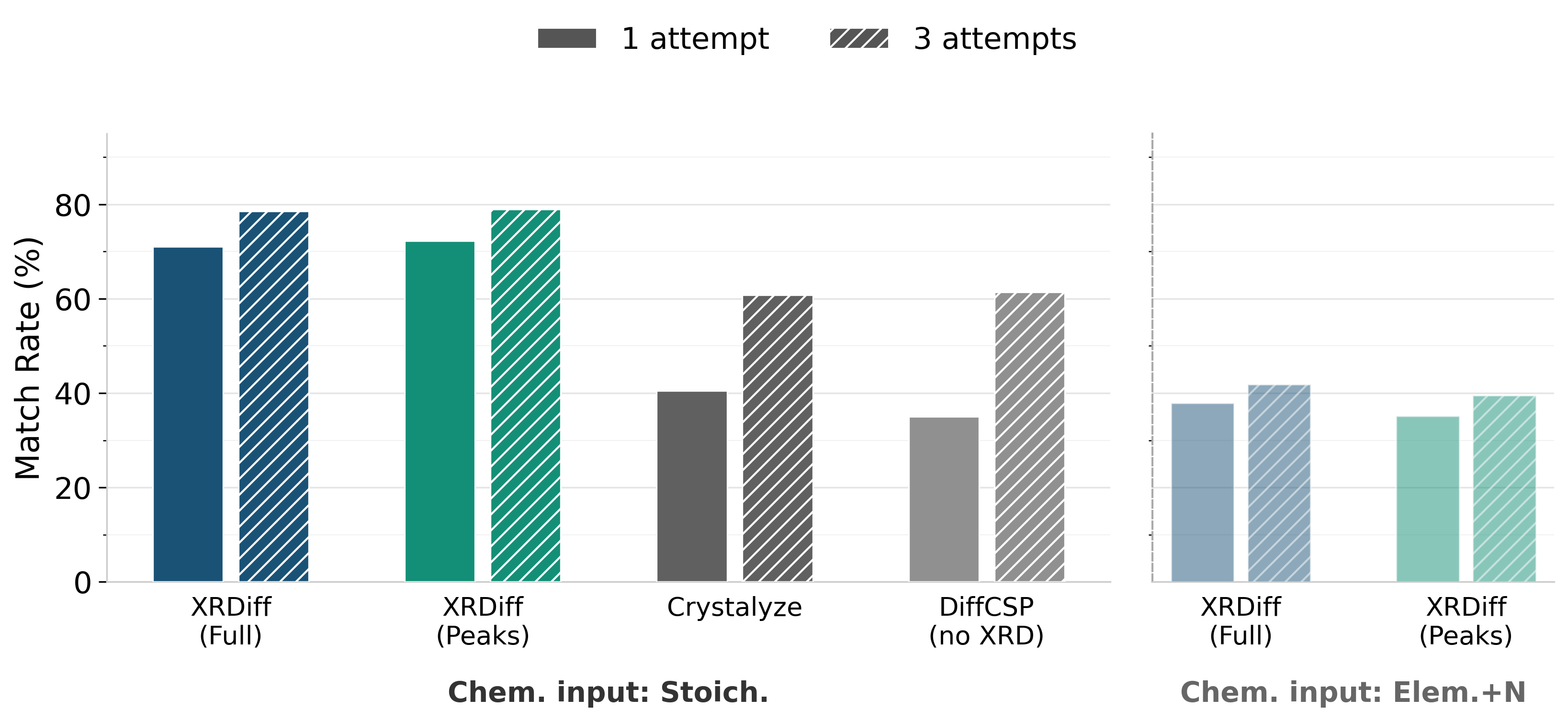}
        \caption{OQMD-Poly}
        \label{subfig:mr_oqmd}
    \end{subfigure}

    \caption{Match rate comparison on Alex-MP-20-Poly (top) and OQMD-Poly (bottom).
    Solid and hatched bars correspond to 1 and 3 attempts, respectively.
    Results are shown for XRDiff (Full PXRD and Peaks variants), Crystalyze, and
    DiffCSP (no XRD input), all using stoichiometry as chemistry input (left panels).
    Faded bars (right panels) show XRDiff results under the Elem.\!+\!N
    chemistry input setting, where only element types and atom count are provided.}
    \label{fig:mr}
\end{figure}

Figure~\ref{fig:mr} report the match performance of XRDiff on CSP from PXRD input across the Alex-MP-20-Poly and OQMD-Poly datasets. The data splits are composition-based, meaning all polymorphs of a given composition are assigned to the same split.

XRDiff achieves strong match rates across both datasets and input types. Root mean squared distance (RMSD) scores are consistently low, as can be seen in Table~\ref{tab:RMSD_comparison}. 

Using the full PXRD spectrum, the model recovers the correct structure in 59.0\% and 71.1\% of cases on Alex-MP-20-Poly and OQMD-Poly, respectively (1 attempt), rising to 72.3\% and 78.6\% with 3 attempts.  Using PXRD peak features yields comparable performance overall, particularly on OQMD-Poly, where match rates are nearly identical to the full-spectrum setting, while showing a moderate decrease on Alex-MP-20-Poly. These results suggest that the model can extract most of the structural information required for generation from either representation in clean computational data.

The Earth Mover's Distance (EMD) results in Table~\ref{appndx:amd} further support the similarity between the full-spectrum and PXRD-peak representations. As a metric based on the distribution of interatomic distances, EMD measures structural similarity in real space, with lower values corresponding to structures that are geometrically closer to the ground truth. Across both datasets, the two input types yield nearly identical EMD values, indicating comparable geometric fidelity of the generated structures. 

To verify that the model genuinely uses the PXRD input rather than generating plausible structures for a given composition, we include a no-PXRD baseline. The large drop in match rate and increase in EMD without PXRD conditioning confirm that the model relies on the diffraction signal to determine the output geometry. The improvement from 1 to 3 attempts indicates stochastic polymorph sampling.

Under the partial composition (denoted Elem.+N in Figures~\ref{fig:mr},~\ref{fig:rruff_mr}) setting, where the full stoichiometry is replaced by atom types and unit cell count only, match rates drop to 28.7\% and 37.9\% on Alex-MP-20-Poly and OQMD-Poly, respectively (1 attempt, full PXRD), rising to 37.5\% and 41.8\% with 3 attempts. The performance gap relative to the full stoichiometry setting is expected, as stoichiometry provides a strong prior that constrains the generation space.

\subsubsection{Lattice Evaluation} \label{sec:lattice_eval}
Figure~\ref{fig:lat_par} visualizes the accuracy of the generated lattice parameters on the Alex-MP-20-Poly and OQMD-Poly test sets through parity plots of the predicted lattice lengths for XRDiff conditioned on the full PXRD representation and full stoichiometry. In both datasets, the majority of predictions are concentrated around the parity line, indicating generally accurate lattice generation. To visualize both the dense concentration of near-perfect predictions and the substantially lower-density regions corresponding to larger prediction deviations, point density is displayed on a logarithmic scale. 

Table~\ref{tab:lattice_mae_comparison} compares the mean absolute error (MAE) of the predicted lattice parameters between XRDiff conditioned on the two PXRD representation variants and the baseline Crystalyze model, all trained with full stoichiometry. Within XRDiff, conditioning on peak features yields consistently lower MAE values than conditioning on the full PXRD representation. However, this improvement is primarily driven by a small number of large-error outliers in the full-PXRD setting rather than by a systematic difference in the overall prediction distribution. In particular, the full-PXRD representation occasionally produces unrealistically large lattice lengths, especially for the $c$ lattice parameter, which substantially inflates the reported MAE despite the strong concentration of predictions near the parity line.

This interpretation is supported by the parity plots of XRDiff conditioned on the full-PXRD representation and on peak features (see SI, Figure~\ref{appndx_fig:lat_par}), which exhibit very similar distributions around the parity line across both datasets. Consistent with this observation, both representations achieve comparable structure match rates (Figure~\ref{fig:mr}), indicating that the overall lattice prediction quality remains similar despite the difference in MAE.

Crystalyze achieves lattice-parameter MAE values comparable to, and in some cases lower than, those of XRDiff, particularly for the $c$ lattice parameter. This behavior is expected given the Crystalyze architecture, which builds upon CDVAE \cite{xie2021crystal}, and is constructed of a Variational Autoencoder in which dedicated multilayer perceptrons are trained directly on the latent representation to predict lattice parameters, atom types, and atom counts.

Additionally, Figure~\ref{appndx_fig:elemn_lat_par} and Table~\ref{tab:lattice_mae_comparison} present results for the accuracy of generated lattice parameters by XRDiff trained with partial composition input. The parity plots show a scatter distribution closely resembling the stoichiometry-conditioned setting, yet the MAE is larger due to further inflated outliers. We therefore attribute the lower match rate in the Elem.+N setting primarily to the model's difficulty in learning correct stoichiometric ratios from partial composition, as well as finer-grained errors in atomic coordinates, rather than a systematic degradation in lattice parameter prediction.

\subsubsection{Match Rate Analysis by Structural Properties} \label{sec:mr_analysis}
\begin{figure}[ht]
    \centering
    \begin{subfigure}[t]{0.98\textwidth}
        \centering
        \includegraphics[width=\linewidth]{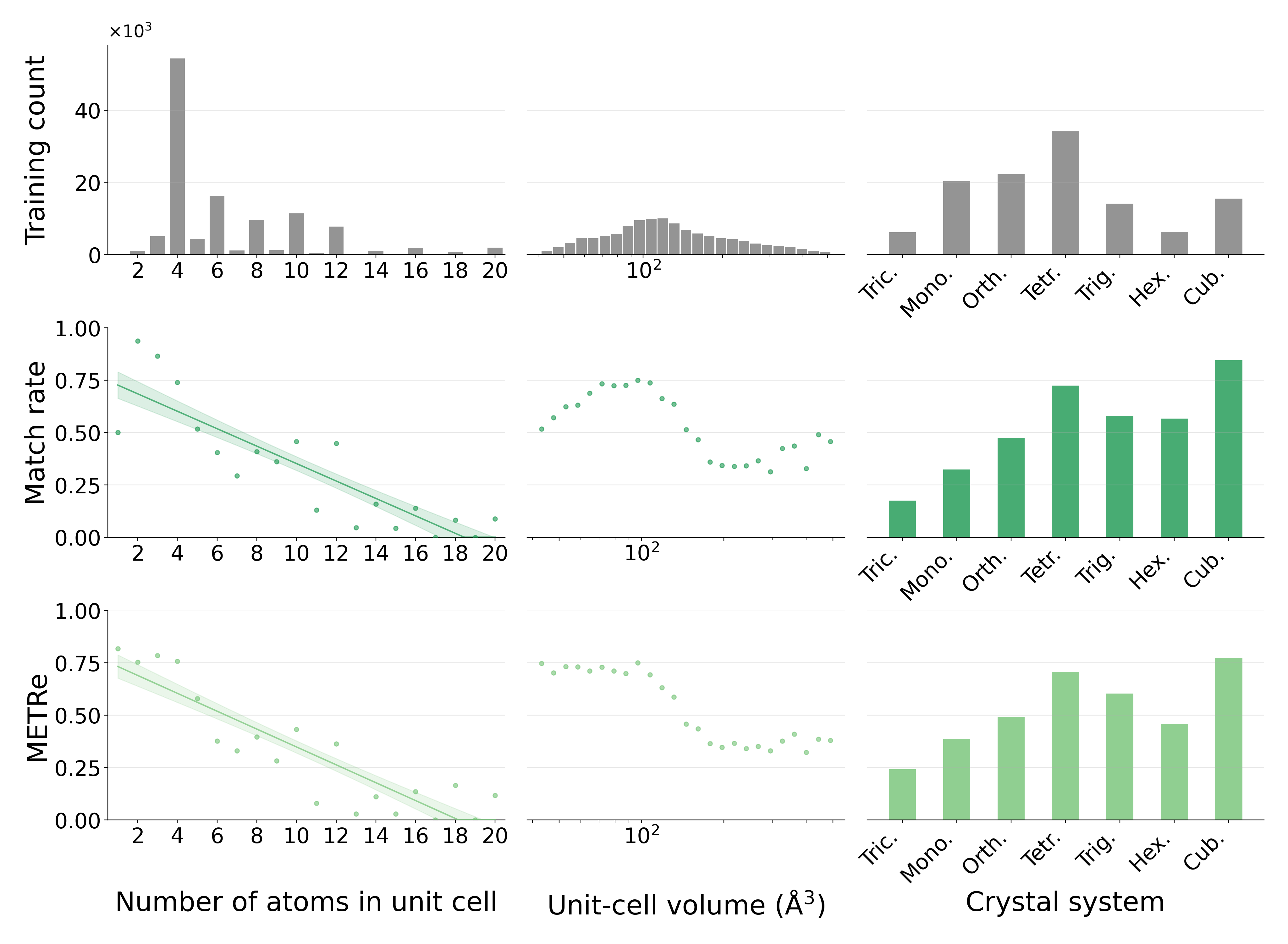}
    \end{subfigure}  

    \caption{\textbf{Analysis of XRDiff performance on Alex-MP-20-Poly as a function of number of atoms in the unit cell, unit-cell volume, and crystal system.} \textbf{Top}: number of training structures in each attribute bin. \textbf{Middle}: corresponding XRDiff test-set match rates, with linear fits shown where applicable. \textbf{Bottom}: DiffCSP (diffusion backbone without PXRD input) test-set METRe rate. Similar trends are observed for both models, with performance decreasing for structures containing more atoms in the unit cell and generally improving for crystal systems with higher symmetry.}
    \label{fig:fails}
\end{figure}

Figure~\ref{fig:fails} analyzes XRDiff performance on Alex-MP-20-Poly as a function of three structural attributes: number of atoms in the unit cell, unit-cell volume, and crystal system. All results in this analysis use full stoichiometry conditioning. The top row shows the number of training structures in each attribute bin. The middle row reports XRDiff test-set match rate within each bin, computed as the fraction of structures whose predicted reconstruction is accepted by \texttt{StructureMatcher} with default settings. Match rates are calculated independently per bin and therefore do not correspond to the overall dataset match rate. Red lines indicate least-squares linear fits where applicable, with shaded regions denoting $\pm1$ standard error of the estimated regression line. The bottom row reports analogous results for DiffCSP, the diffusion backbone used by XRDiff without PXRD conditioning, evaluated using the METRe metric \cite{martirossyan2025all}. Since DiffCSP does not receive PXRD input and therefore is not conditioned on a target polymorph, METRe counts a generated structure as a match if it is structurally equivalent to any polymorph in the test set. Figure~\ref{fig:appnds:fails} presents the corresponding analysis for OQMD-Poly.

With respect to the number of atoms in the unit cell, the match rate trends downwards as the number of atoms increases. This trend is also correlated with training-set representativeness. For the unit-cell volume, both datasets exhibit a non-monotonic dependence: small-volume structures initially yield moderate match rates, followed by a peak at intermediate volumes, and a subsequent decline for larger volumes. In addition, increasingly symmetric crystal systems \cite{hahn1983international, klein2026mineral, dentglasser2001symmetry} consistently exhibit higher match rates. 

The breakdown of matched structures for DiffCSP reveals consistent trends with those observed with XRD signal. In particular, the dependence on structural complexity, both in terms of symmetry and number of atoms, remains unchanged. This consistency suggests that the observed limitations are not specific to the PXRD-to-structure mapping problem, but instead reflect intrinsic challenges in the generation process.

\begin{figure}[ht]
    \centering
    \begin{subfigure}[t]{0.99\textwidth}
        \centering
        \includegraphics[width=\linewidth]{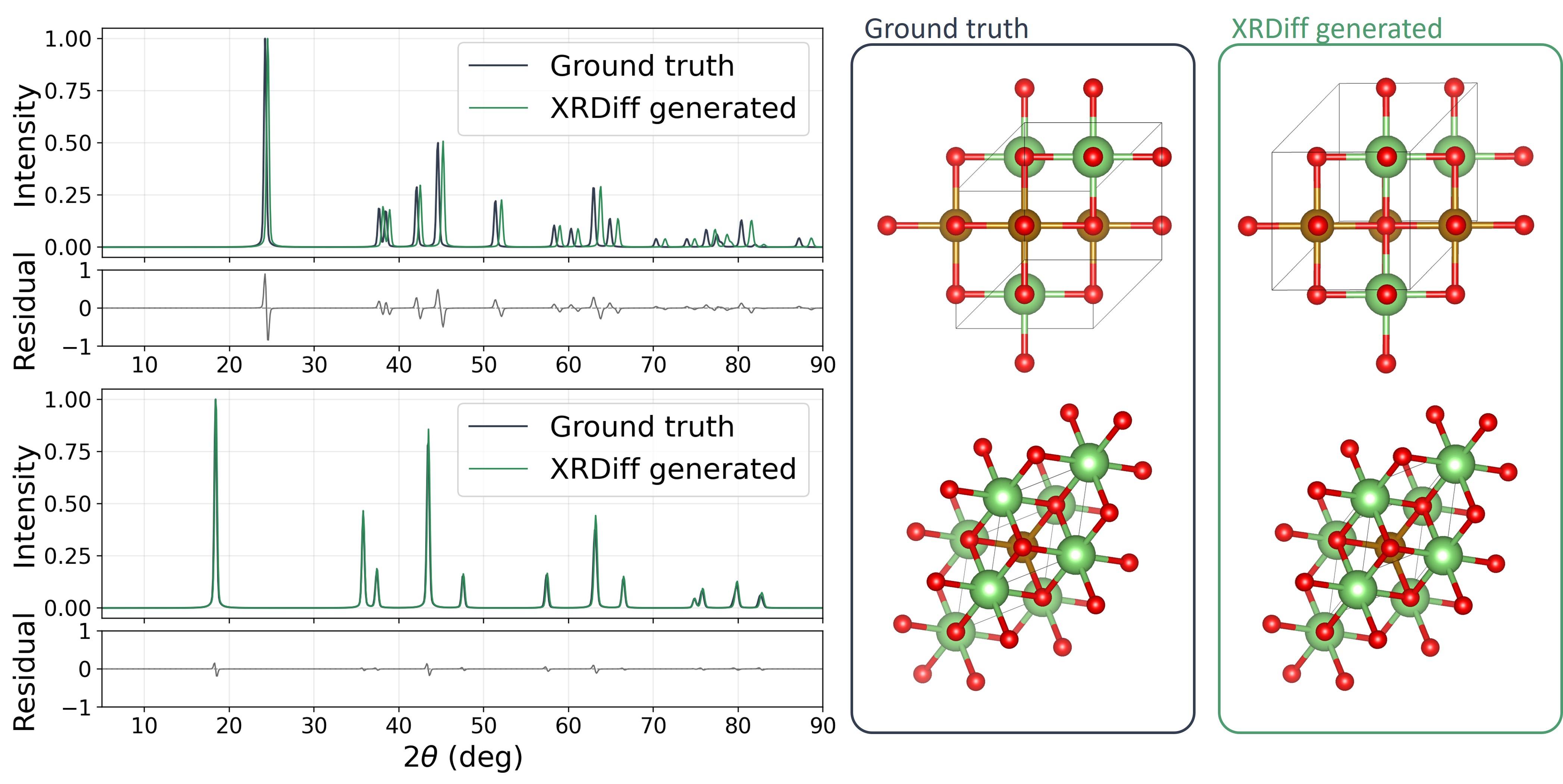}
        \caption{\ce{LiFeO2}}
        \label{subfig:lifeo2}
    \end{subfigure}
    
    \begin{subfigure}[t]{0.99\textwidth}
        \centering
        \includegraphics[width=\linewidth]{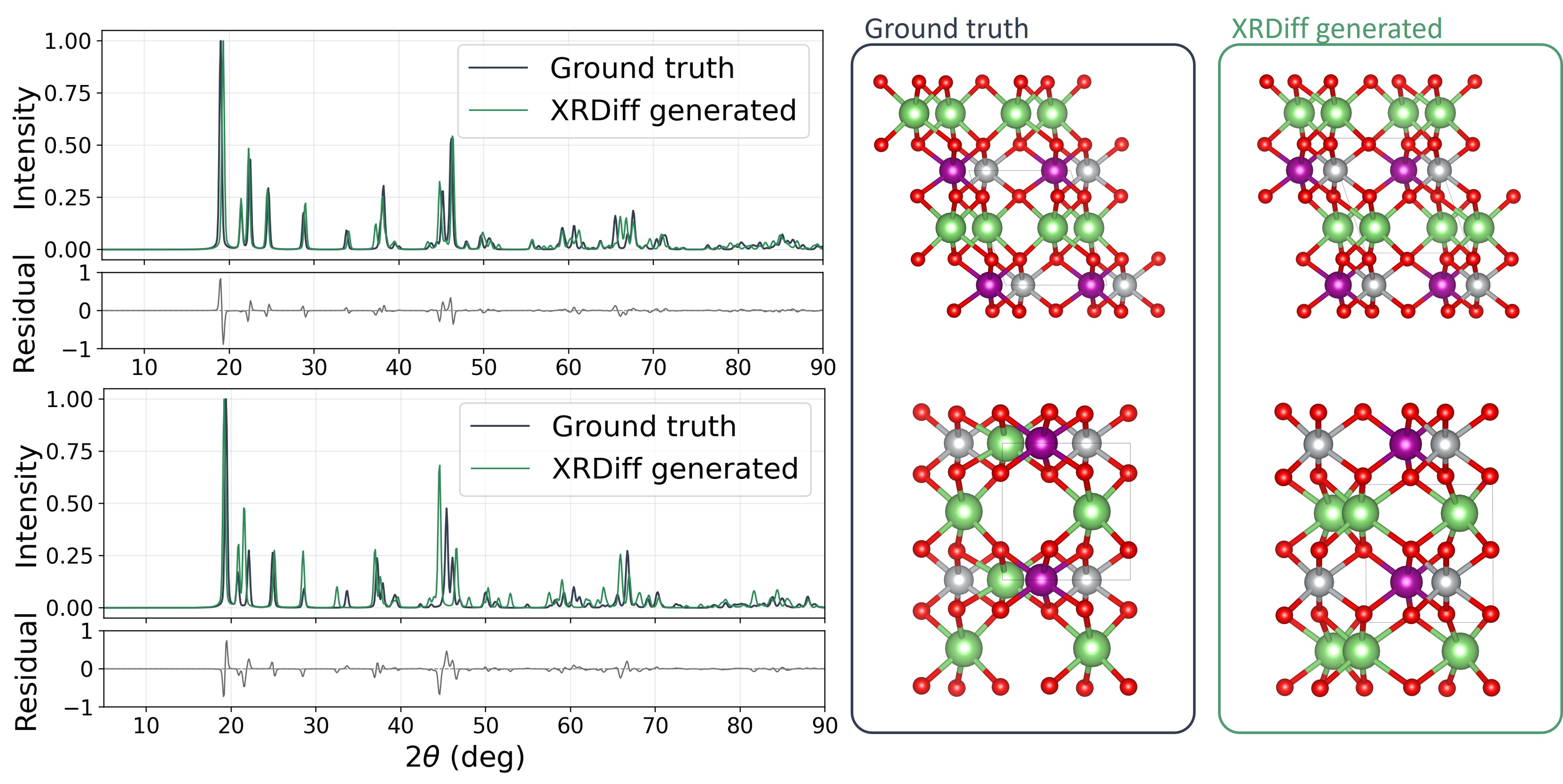}
        \caption{\ce{Li2MnNiO6}}
        \label{subfig:li2mnnio6}
    \end{subfigure}
    
    \caption{\textbf{Representative polymorph pairs from the Li-Fe-Ni-Mn-O chemical space.} Each panel compares ground-truth structures from the OQMD test split with their corresponding XRDiff generations and the overlaid PXRD patterns. Despite the high similarity between diffraction signatures of different polymorphs, XRDiff captures subtle spectral differences and reconstructs distinct crystal geometries.}
    \label{fig:polymorphs_Li-Fe-Ni-Mn-O}
\end{figure}

\subsubsection{Qualitative Analysis} \label{sec:qualitative}
Figure~\ref{fig:polymorphs_Li-Fe-Ni-Mn-O} presents representative randomly selected polymorph pairs from the Li-Fe-Ni-Mn-O chemical space in the OQMD test split together with their corresponding XRDiff generations. 

We note that the diffusion process is invariant to orthogonal transformations of the lattice and periodic translations of the fractional coordinates. Consequently, generated structures may appear rotated or translated relative to the ground-truth structures without indicating an error in the predicted crystal geometry.

The examples include generated polymorphs of \ce{LiFeO2} (Figure~\ref{subfig:lifeo2}) and \ce{Li2MnNiO6} (Figure~\ref{subfig:li2mnnio6}). For \ce{LiFeO2}, XRDiff accurately reproduces the overall crystal geometry and generates diffraction patterns in strong agreement with the targets. In the top example, a small shift in peak positions is visible, corresponding to a minor mismatch in the predicted lattice parameters; however, the relative peak intensities and overall spectral profile remain highly consistent.

For \ce{Li2MnNiO6}, the generated structures similarly recover the main structural motifs and diffraction signatures of the target polymorphs. Peak positions align closely, indicating accurate prediction of lattice parameters, while small local deviations in atomic arrangement remain visible. In the top example, the \ce{Ni-Mn} layer is slightly shifted relative to the ground truth, whereas in the bottom example a \ce{Li} atom is displaced by approximately one layer. These local perturbations lead to minor intensity differences despite the close agreement in peak locations.

Importantly, the displayed polymorph pairs exhibit highly similar diffraction signatures, making phase discrimination challenging. XRDiff nevertheless captures these subtle spectral differences and translates them into distinct crystal geometries, demonstrating sensitivity to fine structural variations encoded in the diffraction signal.

\subsection{Evaluation on Experimental Data} \label{sec:exp_results}

\begin{figure}[H]
    \centering
    \begin{subfigure}[t]{0.99\textwidth}
        \centering
        \includegraphics[width=\linewidth]{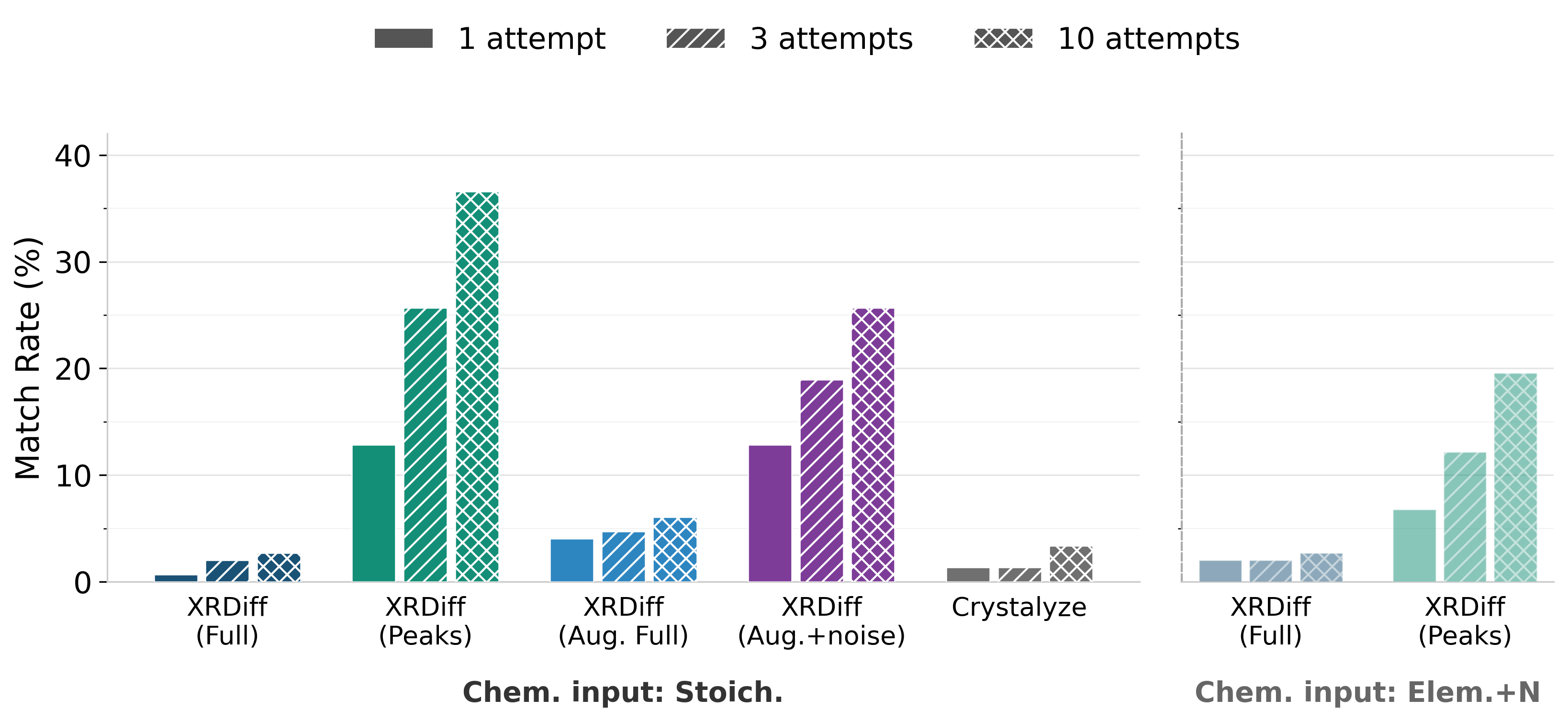}
        \caption{Trained on Alex-MP-20-poly}
        \label{subfig:mr_rruff_alex}
    \end{subfigure}
    
    \begin{subfigure}[t]{0.99\textwidth}
        \centering
        \includegraphics[width=\linewidth]{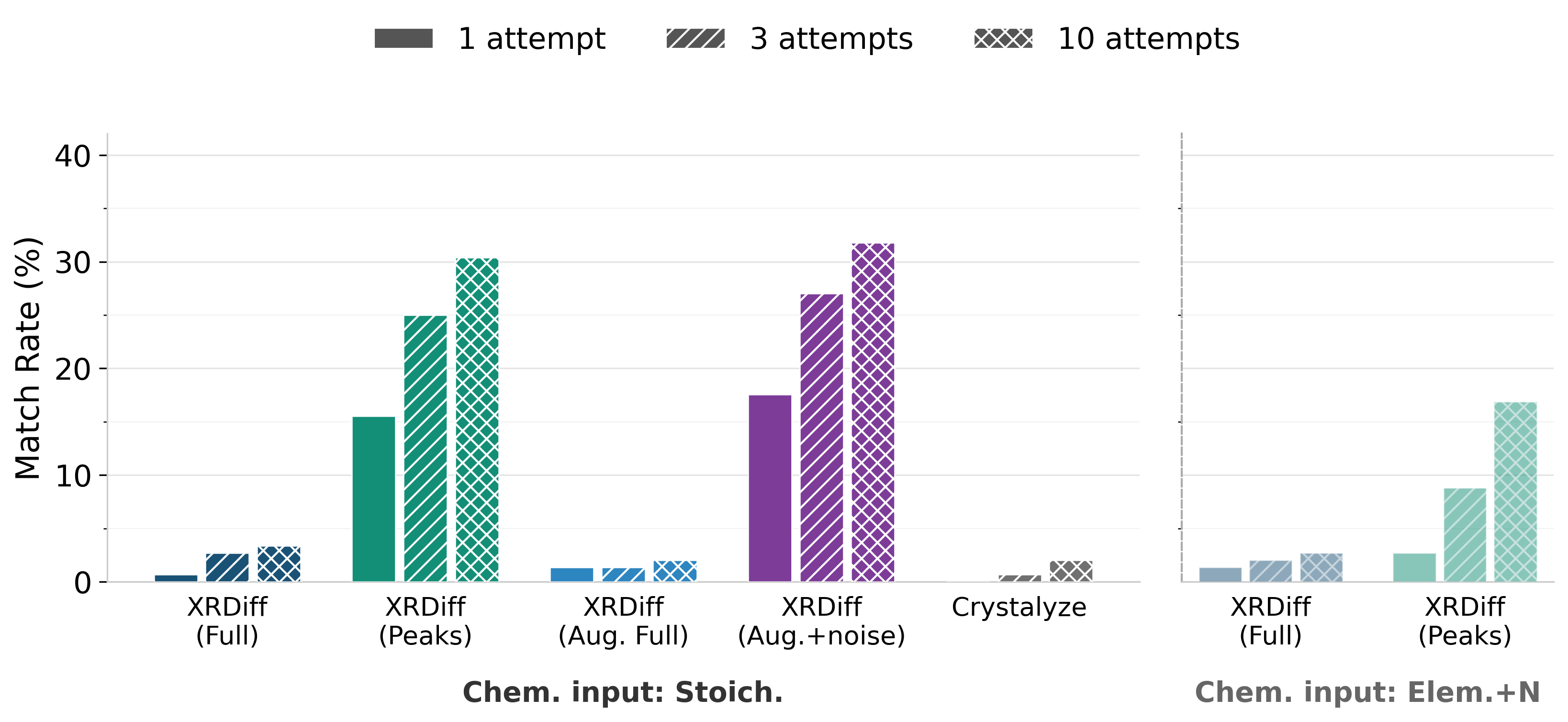}
        \caption{Trained on OQMD-poly}
        \label{subfig:mr_rruff_oqmd}
    \end{subfigure}

    \caption{Match rate on the RRUFF experimental subset, trained on Alex-MP-20-Poly
    (top) and OQMD-Poly (bottom). Solid, hatched and cross-hatched bars correspond to 1, 3 and 10 attempts, respectively. Results are shown for XRDiff under four PXRD input variants (Full, Peaks, Augmented Full, and Augmented Full with noise) and for Crystalyze, all using stoichiometry as chemistry input (left panels). Faded bars (right panels) show XRDiff results under the Elem.\!+\!N chemistry input setting, where only element types and atom count are provided.}
    \label{fig:rruff_mr}
\end{figure}

Figure~\ref{fig:rruff_mr} reports zero-shot performance on the RRUFF experimental benchmark (Section~\ref{sec:rruff_data}), using the same PXRD and stoichiometry inputs as in the computational setting.

Using the full PXRD spectrum, performance drops sharply relative to computational benchmarks, with match rates below 1\% at a single attempt across both training datasets. Increasing the number of attempts to 3 and 10 leads to only marginal improvements. This limited gain suggests that full-spectrum representations remain sensitive to experimental measurement artifacts that are absent from simulated PXRD data, restricting generalization across the simulation-to-experiment gap.

In contrast, the peak-based representation generalizes substantially better to experimental data, achieving match rates of 12.84\%, 25.67\%, and 36.56\% when trained on Alex-MP-20-Poly, and 15.54\%, 25.00\%, and 30.40\% when trained on OQMD-Poly, for 1, 3, and 10 attempts, respectively. By compressing the PXRD pattern into peak-level descriptors, including peak position, intensity, prominence, and width, this representation removes much of the measurement-dependent signal while preserving structurally informative information.

Additionally, performance is evaluated for models trained on heavily augmented computational spectra, a common strategy for improving robustness to experimental PXRD data. The augmentation procedure, described in Section~\ref{sec:aug_data}, can be broadly divided into two groups: (i) modifications to peak appearance, including peak broadening and simulation of preferred orientation through peak reweighting, and (ii) dynamic background noise added during training, with preferred orientation and background noise fitted to the experimental data distribution. 

Results show that augmentations affecting only peak appearance provide limited improvements over training on unmodified spectra, whereas introducing background noise leads to substantial gains in match rate. Under OQMD-Poly training, performance obtained using the full augmentation pipeline becomes largely comparable to that achieved using the proposed peak-based representation, with differences corresponding to only approximately one to two additional correctly matched structures. Under Alex-MP-20-Poly training, however, the peak-based representation achieves larger improvements as the number of attempts increases and attains higher overall match rates. While achieving similar performance, augmentation-based approaches introduce substantial overhead in augmentation design, parameter fitting, and training time.

Under the relaxed Elem.+N setting, the peak-based representation again outperforms the full spectrum. As expected, performance is lower than under full stoichiometry conditioning, reflecting the additional difficulty of the task.

These results suggest that closing the simulation-to-experiment gap can be achieved through input representations that are inherently robust to measurement artifacts, without requiring increasingly complex simulation and augmentation pipelines.

\section{Conclusions}\label{discussion}
The strong match rates achieved by XRDiff demonstrate that a diffusion model conditioned on PXRD learns a meaningful mapping from diffraction patterns to crystal structures. This mapping is sensitive enough to discriminate structurally distinct polymorphs of the same composition, and general enough to hold across compositions and structure types.

Breaking down match rates by structural properties reveals that performance degrades 
with increasing structural complexity, both in terms of the number of atoms in the 
unit cell and decreasing crystal symmetry. The 
observation that the diffusion backbone without PXRD input exhibits consistent 
trends with XRDiff suggests that at least part of the performance gap reflects 
challenges inherent to the generative process and data coverage, rather than the 
PXRD-to-structure mapping specifically. This points to improvements in generative 
modeling and increased data coverage, particularly for low-symmetry structures and 
large unit cells, as promising directions for closing this gap.

The Elem.+N setting more closely reflects realistic experimental scenarios, where 
the precise stoichiometry of an unknown phase may not be available and the 
practitioner works from a known element set. In principle, sampling or scanning over possible values of $N$ at 
inference time could allow the model to approximate a posterior over crystal 
structures consistent with the observed diffraction pattern, more faithfully 
mimicking the exploratory nature of experimental structure determination. 
Performance is nonetheless lower than under full stoichiometry conditioning, 
as the model must resolve ambiguity in atomic ratios that are only weakly encoded 
in the diffraction signal. Exploiting the information carried by relative Bragg 
peak intensities about species occupancy and scattering weight more explicitly 
represents a promising direction for improving performance in this setting.

The simulation-to-experiment gap can be addressed through three broad strategies: data augmentation, noise removal, or input representations that are inherently robust to measurement artifacts. Our results show that among the approaches evaluated, the peak-based representation achieves the strongest generalization to experimental data with the least experimental overhead. Augmentation requires careful design and parameter fitting to the experimental noise distribution and incurs significant training overhead. Noise removal requires prior knowledge of the specific artifacts present. The peak-based representation sidesteps both issues: by discarding the artifact-contaminated background and retaining only peak-level descriptors, it achieves strong generalization without experiment-specific design choices, at least for the class of artifacts present in the RRUFF benchmark. Artifacts that directly affect peak-level properties, such as severe preferred orientation or peak overlap, remain a potential limitation and motivate further investigation of representations that are robust to a broader range of experimental conditions.

\section{Limitations}\label{limitations}
XRDiff assumes the chemical composition is at least partially known. In the partial setting, the model requires the set of atom types and the number of atoms in the unit cell. While the atom types are often known from synthesis conditions, the number of atoms in the unit cell is generally not directly accessible from experimental information alone. The most realistic laboratory scenario is therefore one where only the atom types and possibly flexible ratios are known. Extending the model to operate under this setting, for instance, by sampling candidate unit cell sizes from a learned distribution, is an important direction for future work.

The model also produces multiple candidate structures without ranking them by likelihood or providing uncertainty estimates. In practice, identifying the most plausible candidate among several attempts is important, and incorporating confidence estimation is a valuable direction for future work.

XRDiff assumes a single pristine crystalline phase. Real laboratory samples frequently contain mixtures of phases, and handling such cases requires both detecting the presence of a mixture and decomposing the observed pattern into contributions from individual phases. Extending structure determination methods to multi-phase samples remains an important open problem.

\section{Methods}\label{method}
\subsection{Datasets} \label{subsec:dataset}
\subsubsection{Computational Data}
We conduct experiments on two computational crystal structure datasets: Alex-MP-20 \cite{zeni2023mattergen}, and OQMD v1.5 \cite{saal2013materials}. Alex-MP-20 combines structures from the Alexandria and Materials Project databases, filtered to thermodynamically stable materials ($E_{hull}<0.1 eV$), while OQMD v1.5 is an ICSD-derived database that also covers metastable and hypothetical materials.

Our goal is to generate crystal structures conditioned on PXRD patterns. To encourage the model to utilize the PXRD signal meaningfully, we focus on compositions exhibiting polymorphism, i.e., those associated with multiple distinct crystal structures. In datasets dominated by unique compositions, the mapping from composition to structure can be effectively one-to-one, which may allow the model to largely ignore the PXRD condition. Our setting forces the model to rely on PXRD to correctly disambiguate between structurally distinct polymorphs. Concretely, we remove compositions that appear only once in each dataset and retain only those with at least two structurally distinct entries. We refer to the resulting subsets of Alex-MP-20 and OQMD used for training and evaluation as \textbf{Alex-MP-20-Poly} and \textbf{OQMD-Poly}, containing approximately 150k and 510k samples, respectively.

The datasets are partitioned at the composition level to prevent information leakage between training and evaluation. Specifically, all polymorphs corresponding to a given composition are assigned to the same split. We adopt an 80:10:10 split for training, validation, and testing, respectively.

Diffraction peaks are computed using the \texttt{pymatgen} \texttt{PXRDCalculator}, which returns peak intensities and positions as delta functions. Peak profiles are then modeled using the Pseudo-Voigt approximation, which represents each peak as a linear combination of Gaussian and Lorentzian components:
\begin{equation*}
pV(x) = \eta  G(x) + (1 - \eta) L(x)
\end{equation*}
where $G(x)$ and $L(x)$ are the Gaussian and Lorentzian functions, respectively, and $\eta \in [0,1]$ is the mixing parameter controlling their relative contributions. Further details on the peak profile construction are provided in Supplementary Information~\ref{appndx:PXRD_rep}, and the full PXRD conditioning pipeline is described in Section~\ref{subsec:PXRDcond}.

\subsubsection{Experimental Data} \label{sec:rruff_data}
We evaluate on a curated subset of the RRUFF database~\cite{lafuente20151}, as used by the baseline~\cite{riesel2024crystal}. The curation process presented in \cite{riesel2024crystal} retained a single diffraction pattern per mineral to avoid over-representation, and each pattern was manually verified to ensure consistency between the reported crystal system and lattice parameters in the structure and PXRD data files. Any training set samples that overlapped with this benchmark were removed from training to prevent data leakage. The final benchmark consists of 148 structures.

\subsubsection{Augmented Data} \label{sec:aug_data}
To compare against augmentation-based approaches for bridging the simulation-to-experiment gap, we construct augmented versions of the training datasets following \cite{riesel2024crystal}. Augmentations are divided into two groups. The first group modifies peak appearance: spectra are re-simulated with varying Caglioti broadening parameters, and individual peak intensities are perturbed by Gamma-distributed multiplicative factors calibrated to match observed-to-simulated intensity ratios in the RRUFF subset, mimicking preferred orientation and other non-ideal scattering effects. The second group applies dynamic background noise online during training, sampling additive Gamma-distributed noise traces from a bivariate log-normal distribution fitted to background profiles extracted from the RRUFF calibration subset. Full details of both augmentation procedures are provided in the SI (Section~\ref{appndx:aug_data}).

Unlike the proposed peak-features representation, these augmentations require empirical calibration using experimental diffraction data and substantially increase both dataset size and preprocessing and training costs.

\subsection{Diffusion-Based Crystal Structure Prediction}
We build upon the DiffCSP framework \cite{jiao2023crystal}, which formulates crystal structure prediction as a generative task using a score-based diffusion model. Given a chemical composition, the objective is to learn the reverse of a diffusion process that gradually perturbs a crystal’s fractional coordinates $x$ and lattice parameters $L$ with Gaussian noise.

The denoising model is implemented as a graph neural network (GNN), where each crystal is represented as a graph with atoms as nodes and interatomic interactions as edges. Each atom $i$ is associated with a feature vector $\mathbf{h}_i$, and edges encode relative geometric information between neighboring atoms. The network updates node and edge features through multiple message-passing layers, where messages are exchanged along edges and aggregated to refine atomic representations. 

At each diffusion timestep $t$, the model predicts denoised coordinates and lattice parameters from the noisy input, enabling iterative refinement of the crystal structure. In this work, we retain the diffusion process backbone from \cite{jiao2023crystal} and introduce PXRD-based conditioning to guide the generation process, and evaluate the ability to differentiate polymorphs.

\subsection{PXRD Conditioned Diffusion}
In each denoising step $t$, the PXRD spectrum $s$ is encoded and injected into the score network to guide the structural updates.

\subsubsection{PXRD Condition Representation} \label{subsec:PXRDcond}
We consider two complementary representations of the PXRD signal. 

First, the full diffraction pattern is represented as a one-dimensional intensity vector $\mathbf{s} \in \mathbb{R}^{L}$, where $L$ is determined by the discretization resolution of the $2\theta$ axis. The intensity values are normalized to $[0,1]$ and encode the complete diffraction profile. See Section~\ref{subsec:dataset} and  Supplementary Information~\ref{appndx:PXRD_rep} for more details on the PXRD profile calculation.

Second, we construct a peak-based representation that captures salient crystallographic features. Peaks are identified using a standard local maxima detection. For each detected peak $k$, we extract its intensity $I_k$, normalized position $p_k \in [0,1]$, prominence $\pi_k$, and normalized width $w_k$. A fixed-length feature vector is constructed by representing each spectrum with up to 20 highest-intensity peaks. Spectra containing fewer than 20 peaks are zero-padded, while for spectra with more than 20 peaks, only the 20 highest-intensity peaks are retained. For more details on peak featurization, see Supplementary Information~\ref{appndx:PXRD_rep}.

Both representations are encoded using same-depth MLPs to produce latent embeddings of the PXRD signal, which are used in the conditioning mechanism described in Section~\ref{subsec:cond_mech}.

\subsubsection{PXRD Conditioning Mechanism} \label{subsec:cond_mech}
To incorporate the PXRD signal into the generation process, we condition the GNN message-passing layers via feature-wise transformation\textemdash a class of conditioning mechanisms that modulate intermediate representations through learned multiplicative and additive interactions \cite{dumoulin2018feature}.
Specifically, given node or edge features $\mathbf{h} \in \mathbb{R}^d$ within the message passing network, conditioning on an PXRD spectrum $\mathbf{s}$ is implemented via an affine transformation:
\begin{equation}
\mathbf{h} \leftarrow (1 + \boldsymbol{\gamma}(\mathbf{s})) \odot \mathbf{h} + \boldsymbol{\beta}(\mathbf{s}),
\end{equation}
where $\odot$ denotes element-wise multiplication, and 
$\boldsymbol{\gamma}(\mathbf{s}), \boldsymbol{\beta}(\mathbf{s}) \in \mathbb{R}^d$ are learned scaling and shifting functions of the PXRD representation, parameterized by MLPs. This is known as Feature-wise Linear Modulation (FiLM) \cite{perez2018film}.
To stabilize training, all FiLM layers are initialized to perform the identity transformation, i.e., $\boldsymbol{\gamma}(\mathbf{s}) = \mathbf{0}$ and $\boldsymbol{\beta}(\mathbf{s}) = \mathbf{0}$ at initialization. This ensures that the model initially behaves as an unconditional diffusion model, and gradually learns to incorporate PXRD information.

\subsection{Chemical Composition Condition}
We explore two levels of compositional conditioning. In the first, the full stoichiometry is provided: the exact per-atom element assignments are given as input. In the second, only the element set is provided \textemdash a binary indicator of which chemical species are present \textemdash along with the total number of atoms. Under this relaxed condition, the model must infer stoichiometry from the XRD pattern and jointly predict atom types alongside geometry. The type prediction is supervised with cross-entropy against the ground-truth per-atom labels. For more details, see SI Section~\ref{appndx:chem_cond}.

\subsection{Evaluation Metrics}
We evaluate structure reconstruction using both discrete and continuous measures of structural similarity to capture complementary aspects of generation quality. We report the \textbf{match rate}, defined as the fraction of generated structures that exactly match the ground-truth crystal structure. Matches are determined using \texttt{StructureMatcher} with default tolerances (angle tolerance 5$^\circ$, site tolerance 0.3\AA, and lattice tolerance 0.2). \textbf{RMSD} is calculated between the ground truth and the best matching candidate, normalized by $\sqrt[3]{V/N}$ where V is the volume of the lattice, and averaged over the matched structures. 

We additionally report the \textbf{Earth Mover’s Distance (EMD)} between the Pointwise Distance Distribution (PDD) representations of each ground-truth crystal structure and its spectrum-conditioned reconstruction \cite{widdowson2022resolving}. The PDD encodes the local geometric environment of a periodic crystal by constructing, for each atom in the unit cell, an ordered vector of distances to its nearest neighbors in the infinite periodic structure. These vectors are collected into a weighted matrix representation of the crystal structure. Structural similarity between two crystals is then quantified using the EMD between their corresponding PDD representations, providing a continuous geometry-based measure of structural similarity even when an exact crystallographic match is not achieved. This metric is limited, however, as it quantifies structural similarity based solely on geometry and does not account for the atomic species occupying each site.

We evaluate lattice reconstruction accuracy using the \textbf{MAE} of the predicted lattice lengths ($a$, $b$, and $c$) with respect to the ground-truth structure.

\clearpage
\section{Code and Data Availability}
The code and data used in this work are publicly available at \url{https://github.com/learningmatter-mit/xrdiff}, including scripts to reproduce the main results.

\section{Acknowledgments}
N.S.'s research is sponsored by the Eli and Dorothy Berman Fellowship and the National Science Foundation (NSF) under Award Number 2209892 ‘Garden: A FAIR Framework for Publishing and Applying AI Models for Translational Research in Science, Engineering, Education, and Industry’.


\clearpage
\bibliography{sn-bibliography}

\clearpage
\section*{Supplementary Information}

\noindent\textbf{Supplementary Information for:}

\textit{XRDiff: Crystal Structure Prediction from Powder X-Ray Diffraction Data using Diffusion Models}
\vspace{15pt}

\noindent\textbf{Authors:} Nofit Segal, Mingda Li, Benjamin Kurt Miller, Rafael Gomez Bombarelli

\begin{center}
\rule{0.4\textwidth}{0.3pt}
\end{center}

\begin{appendices}
\setcounter{figure}{0}
\renewcommand{\thefigure}{S\arabic{figure}}

\setcounter{table}{0}
\renewcommand{\thetable}{S\arabic{table}}

\setcounter{equation}{0}
\renewcommand{\theequation}{S\arabic{equation}}

\section{PXRD Representation}\label{appndx:PXRD_rep}
For a peak centered at $2\theta_0$, the Gaussian and Lorentzian components are given by:
\begin{align*}
    G(x) &= \exp\left[ -\frac{4 \ln 2 \, (x - 2\theta_0)^2}{H_G^2} \right], \\
    L(x) &= \frac{1}{1 + \frac{4 (x - 2\theta_0)^2}{H_L^2}},
\end{align*}
where $H_G$ and $H_L$ are the full widths at half maximum (FWHM) for the Gaussian and Lorentzian profiles, respectively.

In practice, peak broadening in PXRD is described by the Caglioti relation:
\begin{equation*}
    H^2(2\theta) = U \, \tan^2\theta + V \, \tan\theta + W,
\end{equation*}
where $U$, $V$, and $W$ are the Caglioti parameters. This equation gives the squared FWHM as a function of diffraction angle, and is applied separately for the Gaussian and Lorentzian widths, i.e., $H_G(2\theta)$ and $H_L(2\theta)$. The parameters account for instrumental and sample-dependent broadening effects.

We compute the PXRD profile pattern by summing $pV(x)$ contributions from all Bragg reflections over $2\theta \in [5^\circ, 90^\circ]$, and then discretize the intensity into bins of width $0.1^\circ$. We adopt Caglioti parameterss $U=0.1 , V=0.01, W=0.1$ and $\eta=0.1$. This produces a fixed-length $\text{PXRD}$ vector, $\mathbf{x}$, of size $850$ for each structure.

From this representation, either this is the final one that's inputted to the model and embedded to make up the condition, or the peak's features are extracted through the process outlined in~\ref{subsec:peak_ext}.

\subsection{Peak Features Extraction} \label{subsec:peak_ext}
Peak-level features are extracted from PXRD patterns using \texttt{find\_peaks} function from \texttt{scipy.signal}. Minimum thresholds on height, width, and prominence can constrain peak detection.

For each detected peak at index \( i \), the following quantities are extracted: peak intensity \( x_i \), position \( i \), prominence \( p_i \), and width \( w_i \).

To ensure consistency across samples, peak features are normalized. Peak positions are scaled to the unit interval:
\[
\tilde{p}_i = \frac{i}{N - 1},
\]
where \( N \) is the number of PXRD bins. Peak widths are normalized relative to the maximum width within the same pattern:
\[
\tilde{w}_i = \frac{w_i}{\max_j w_j},
\]
Peak intensities and prominences are preserved in their original scale, noting that all PXRD signals are pre-normalized to unit maximum intensity.

While peak detection itself does not impose an upper bound, the input representation to the model uses a fixed maximum number of peaks. Specifically, peaks are first ranked by prominence, and the top \(K\) peaks (default: \(K = 20\)) are retained.

If fewer than \(K\) peaks are detected, the feature vectors are zero-padded to maintain a consistent input size. Conversely, if more than \(K\) peaks are detected, only the top-\(K\) highest peaks are kept, and the remaining peaks are discarded.

This results in a fixed-size representation per sample, while prioritizing the most structurally significant reflections. Notably, peak selection is based solely on intensity, ensuring robustness to noise and background variations.

\clearpage
\section{Additional Results}
\subsection{Root Mean Squared Distance (RMSD)}
\begin{table}[ht]
\centering
\caption{RMSD $\downarrow$ comparison on Alex-MP-20-Poly and OQMD-Poly datasets.}
\label{tab:RMSD_comparison}
\small
\setlength{\tabcolsep}{4pt}
\renewcommand{\arraystretch}{1.2}
\begin{tabular}{cllccc}
\toprule
\multirow{2}{*}{\textbf{Att.}} &
\multirow{2}{*}{\textbf{Model}} &
\multicolumn{2}{c}{\textbf{Input}} &
\multicolumn{2}{c}{\textbf{RMSD $\downarrow$}} \\
\cmidrule(lr){3-4}
\cmidrule(lr){5-6}
& & \textbf{PXRD} & \textbf{Chem.}
& \textbf{Alex-MP-20-Poly}
& \textbf{OQMD-Poly} \\
\midrule
\multirow{6}{*}{\rotatebox[origin=c]{90}{1 att.}}
& XRDiff & Full   & Stoich. & 0.01 & 0.01 \\
& XRDiff & Peaks  & Stoich. & 0.01 & 0.00 \\
& Crystalyze \cite{riesel2024crystal} & Full     & Stoich. & 0.02 & 0.02 \\
& DiffCSP \cite{jiao2023crystal}      & $\times$ & Stoich. & 0.02 & 0.01 \\
\cdashline{2-6}[0.4pt/1.2pt]
& XRDiff & Full   & Elem.+N & 0.02 & 0.01 \\
& XRDiff & Peaks  & Elem.+N & 0.02 & 0.01 \\
\midrule
\multirow{6}{*}{\rotatebox[origin=c]{90}{3 att.}}
& XRDiff & Full   & Stoich. & 0.01 & 0.01 \\
& XRDiff & Peaks  & Stoich. & 0.01 & 0.01 \\
& Crystalyze \cite{riesel2024crystal} & Full     & Stoich. & 0.02 & 0.02 \\
& DiffCSP \cite{jiao2023crystal}      & $\times$ & Stoich. & 0.07 & 0.01 \\
\cdashline{2-6}[0.4pt/1.2pt]
& XRDiff & Full   & Elem.+N & 0.02 & 0.01 \\
& XRDiff & Peaks  & Elem.+N & 0.02 & 0.01 \\
\bottomrule
\end{tabular}
\end{table}

\subsection{Pointwise Distance Distribution (PDD)} \label{appndx:amd}
\begin{table}[ht]
\centering
\caption{Pointwise Distance Distribution (PDD) $\downarrow$ $\pm$ s.e.m comparison (1 attempt).}
\label{tab:amd_comparison}
\small
\setlength{\tabcolsep}{6pt}
\renewcommand{\arraystretch}{1.4}
\begin{tabular}{l cc}
\toprule
\textbf{PXRD Input} & \textbf{Alex-MP-20-Poly} & \textbf{OQMD-Poly} \\
\midrule
Full PXRD          & 0.250 $\pm$ 0.002 & 0.133 $\pm$ 0.003 \\
PXRD Peaks         & 0.289 $\pm$ 0.002 & 0.139 $\pm$ 0.001 \\
\midrule
\midrule
No PXRD            & 0.568 $\pm$ 0.004 & 0.351 $\pm$ 0.002 \\
\bottomrule
\end{tabular}
\end{table}

\subsection{Lattice Parameters Prediction}

\begin{table}[ht]
\centering
\caption{Mean absolute error (MAE) $\downarrow$ for lattice-parameter prediction ($a$, $b$, and $c$). Values are reported as mean $\pm$ standard error of the mean (s.e.m.)}
\label{tab:lattice_mae_comparison}
\scriptsize
\setlength{\tabcolsep}{5pt}
\renewcommand{\arraystretch}{1.15}
\begin{tabular}{lllccc}
\toprule
\multirow{2}{*}{\textbf{Model}} &
\multirow{2}{*}{\shortstack{\textbf{Chem.}\\\textbf{Input}}} &
\multirow{2}{*}{\shortstack{\textbf{PXRD}\\\textbf{Input}}} &
\multirow{2}{*}{\shortstack{\textbf{Lattice}\\\textbf{param}}} &
\textbf{Alex-MP-20-Poly} &
\textbf{OQMD-Poly} \\
& & & & MAE [\AA] $\downarrow$ & MAE [\AA] $\downarrow$\\
\midrule
\multirow{6}{*}{XRDiff}
& \multirow{6}{*}{Stoich.}
& \multirow{3}{*}{Full PXRD}
& $a$ & 0.62 $\pm$ 0.12 & 0.16 $\pm$ 0.03 \\
& & & $b$ & 0.65 $\pm$ 0.13 & 0.24 $\pm$ 0.04 \\
& & & $c$ & 7.25 $\pm$ 18.8 & 1.76 $\pm$ 0.52 \\
\cmidrule(lr){3-3}
& & \multirow{3}{*}{PXRD Peaks}
& $a$ & 0.23 $\pm$ 0.01 & 0.09 $\pm$ 0.00\\
& & & $b$ & 0.31 $\pm$ 0.01 & 0.11 $\pm$ 0.00 \\
& & & $c$ & 0.80 $\pm$ 0.12 & 0.20 $\pm$ 0.00 \\
\cdashline{1-6}[0.4pt/1.2pt]
\multirow{3}{*}{Crystalyze}
& \multirow{3}{*}{Stoich.}
& \multirow{3}{*}{Full PXRD}
& $a$ & 0.28 $\pm$ 0.00 & 0.23 $\pm$ 0.00 \\
& & & $b$ & 0.28 $\pm$ 0.00 & 0.23 $\pm$ 0.00 \\
& & & $c$ & 0.45 $\pm$ 0.00 & 0.34 $\pm$ 0.01 \\
\midrule
\multirow{6}{*}{XRDiff}
& \multirow{6}{*}{Elem.+N}
& \multirow{3}{*}{Full PXRD}
& $a$ & 0.69 $\pm$ 0.10 & 0.98 $\pm$ 0.11 \\
& & & $b$ & 0.66 $\pm$ 0.09 & 1.48 $\pm$ 0.15 \\
& & & $c$ & 7.24 $\pm$ 1.37 & 14.26 $\pm$ 1.39 \\
\cmidrule(lr){3-3}
& & \multirow{3}{*}{PXRD Peaks}
& $a$ & 0.83 $\pm$ 0.07 & 0.55 $\pm$ 0.05 \\
& & & $b$ & 0.89 $\pm$ 0.07 & 0.63 $\pm$ 0.06 \\
& & & $c$ & 10.23 $\pm$ 1.20 & 11.51 $\pm$ 1.38 \\
\bottomrule
\end{tabular}
\end{table}

\begin{figure}[t]
    \centering
    \begin{subfigure}[t]{0.98\textwidth}
        \centering
        \includegraphics[width=\linewidth]{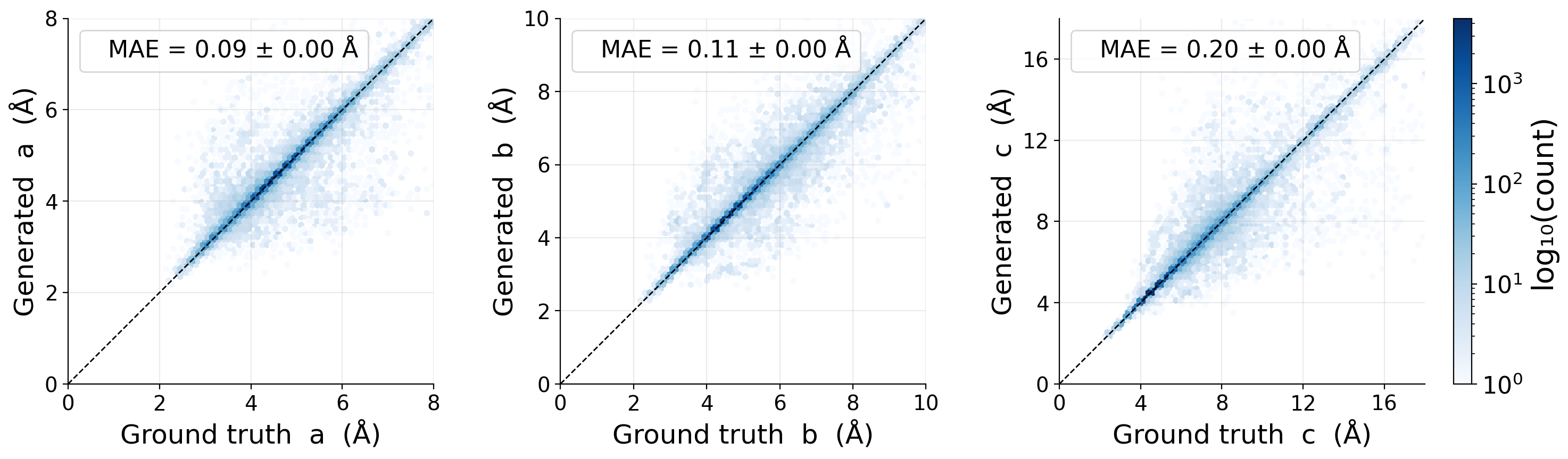}
        \caption{}
        \label{subfig:lat_par_oqmd_peaks}
    \end{subfigure}
    
    \begin{subfigure}[t]{0.98\textwidth}
        \centering
        \includegraphics[width=\linewidth]{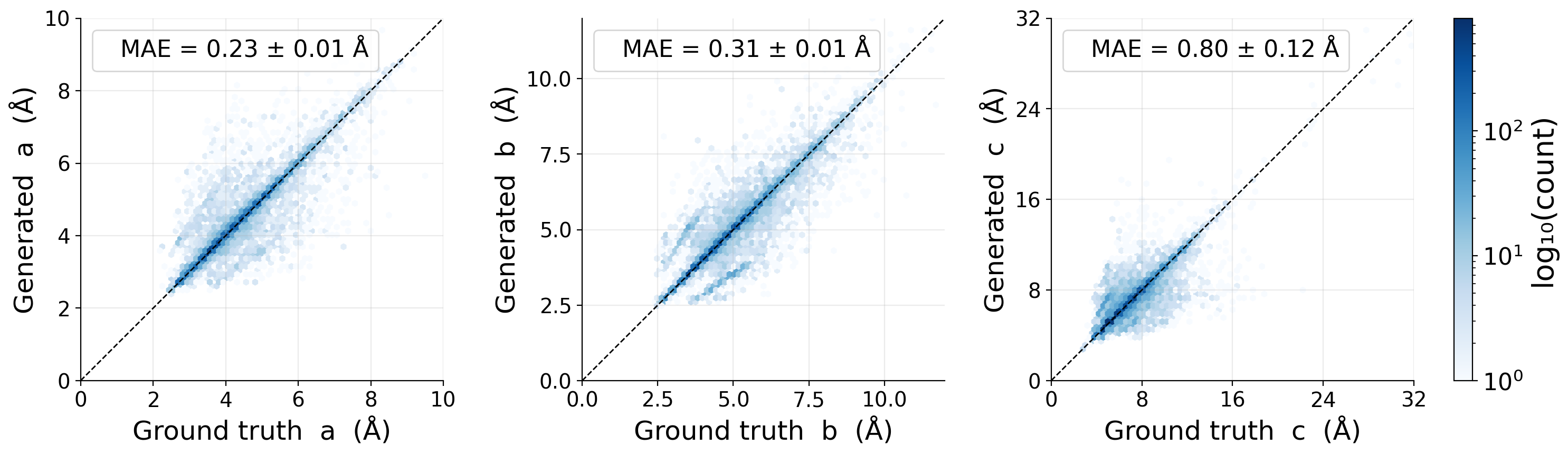}
        \caption{}
        \label{subfig:lat_par_alex_peaks}
    \end{subfigure}

    \caption{\textbf{Parity plots for lattice length prediction by XRDiff on the Alex-MP-20-Poly and OQMD-Poly test sets.} 
    Comparison between ground-truth and generated lattice lengths $a$, $b$, and $c$ for XRDiff conditioned on PXRD \textbf{peaks features and full stoichiometry}. The dashed line denotes perfect agreement. Point colors represent the local scatter density on a logarithmic scale, enabling simultaneous visualization of the highly concentrated regions near the parity line and the substantially lower-density regions associated with larger prediction deviations. Reported values correspond to the mean absolute error (MAE) and the standard error of the mean across the test set.}
    \label{appndx_fig:lat_par}
\end{figure}

\begin{figure}[t]
    \centering
    \begin{subfigure}[t]{0.98\textwidth}
        \centering
        \includegraphics[width=\linewidth]{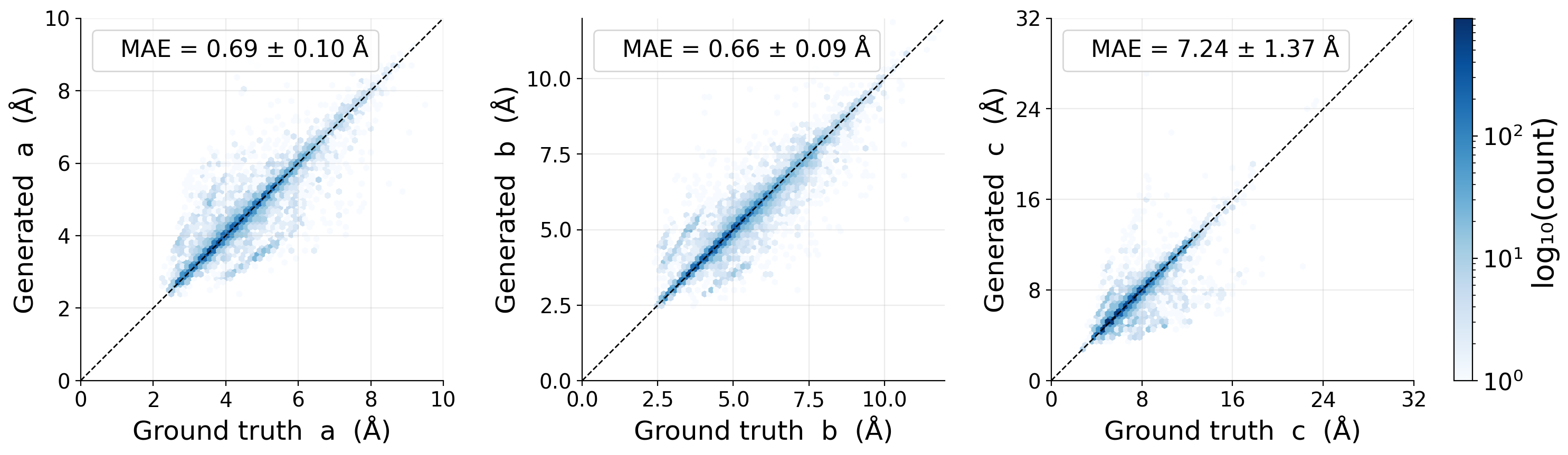}
        \caption{}
        \label{subfig:elenm_lat_par_alex_full}
    \end{subfigure}
    
    \begin{subfigure}[t]{0.98\textwidth}
        \centering
        \includegraphics[width=\linewidth]{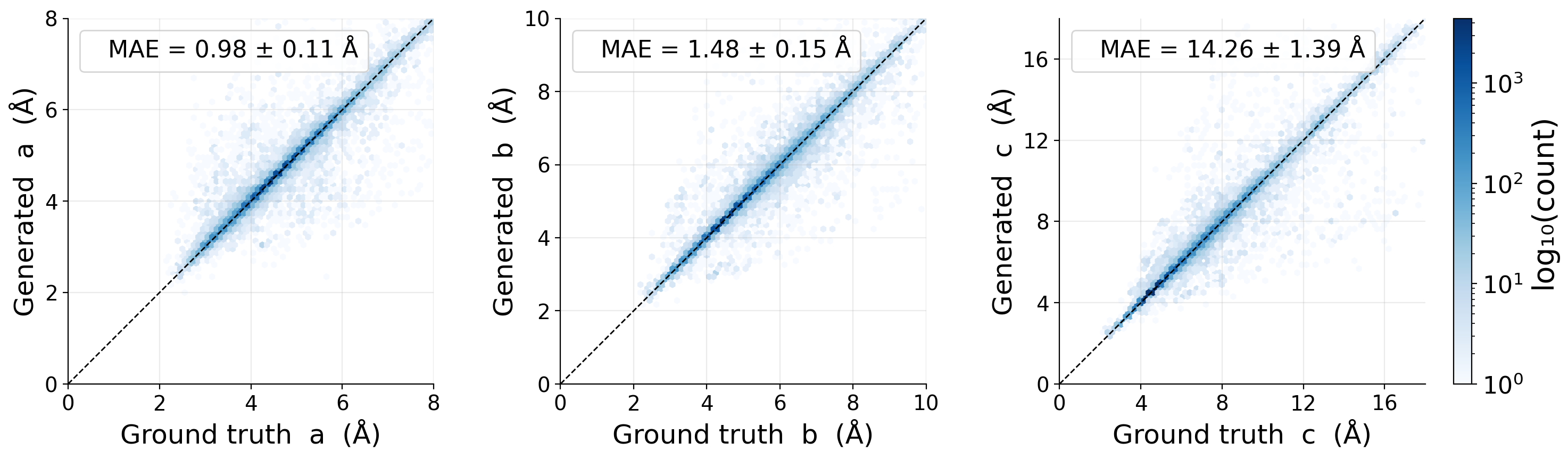}
        \caption{}
        \label{subfig:elemn_lat_par_oqmd_full}
    \end{subfigure}

    \caption{\textbf{Parity plots for lattice length prediction by XRDiff on the Alex-MP-20-Poly and OQMD-Poly test sets.} Comparison between ground-truth and generated lattice lengths $a$, $b$, and $c$ for XRDiff conditioned on \textbf{full PXRD spectra and partial composition} (element types and total number of atoms in the unit cell). The dashed line denotes perfect agreement. Point colors represent the local scatter density on a logarithmic scale, enabling simultaneous visualization of the highly concentrated regions near the parity line and the substantially lower-density regions associated with larger prediction deviations. Reported values correspond to the mean absolute error (MAE) and the standard error of the mean across the test set.}
    \label{appndx_fig:elemn_lat_par}
\end{figure}

\subsection{Matched Structures Analysis}

\begin{figure}[t]
        \centering
            \centering
            \includegraphics[width=\linewidth]{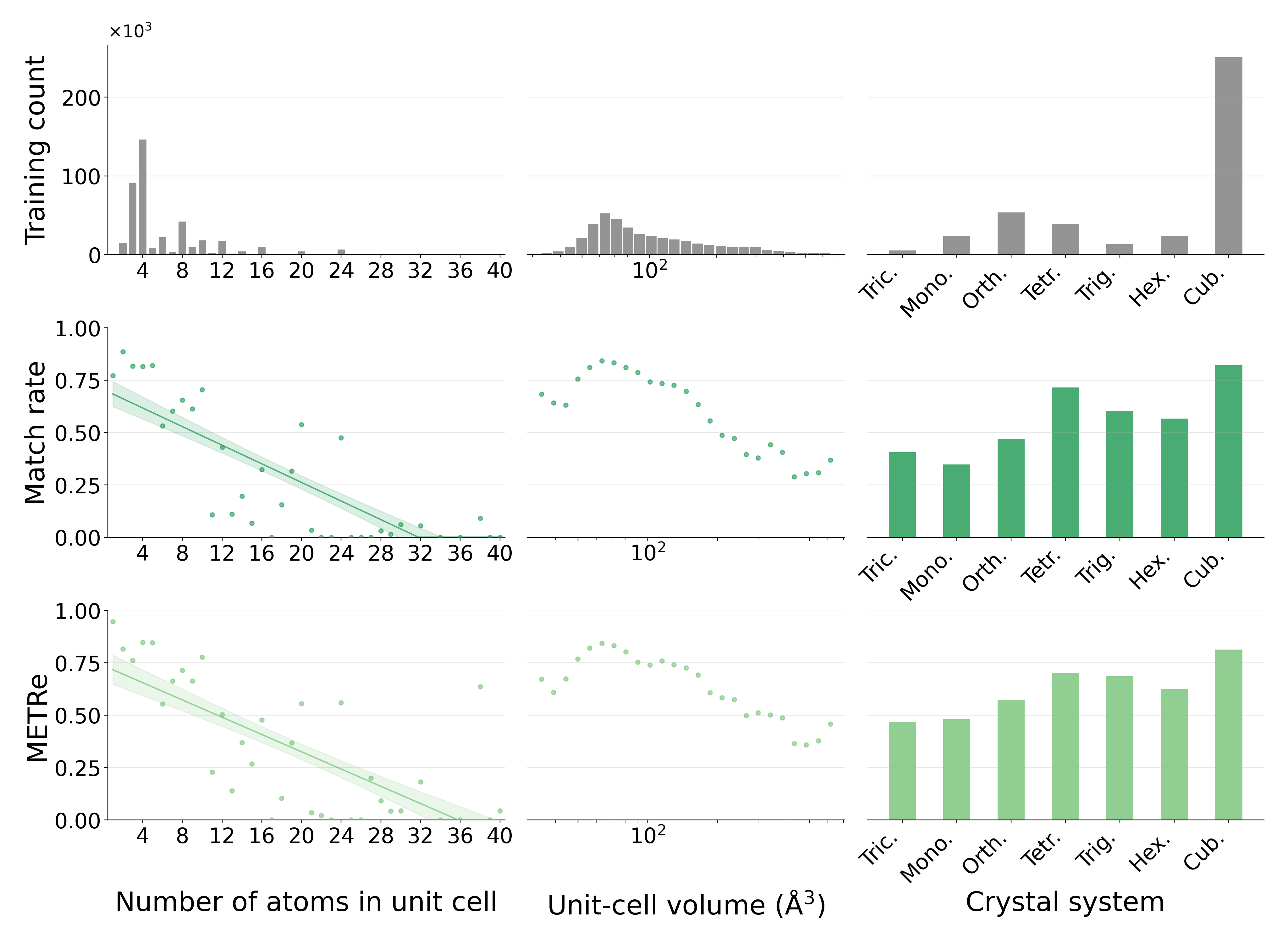}

    \caption{\textbf{Analysis of XRDiff performance on OQMD-Poly as a function of number of atoms in the unit cell, unit-cell volume, and crystal system.} Top: number of training structures in each attribute bin. Middle: corresponding XRDiff test-set match rates, with linear fits shown where applicable. Bottom: DiffCSP (diffusion backbone without PXRD input) test-set METRe rate. Similar trends are observed for both models, with performance decreasing for structures containing more atoms in the unit cell and generally improving for crystal systems with higher symmetry.}
    \label{fig:appnds:fails}
\end{figure}

\clearpage
\section{Implementation Details}
\subsection{Chemical Composition Condition} \label{appndx:chem_cond}
We explore two levels of compositional conditioning:

\textbf{Stoichiometry Conditioning.}
In the full-stoichiometry variant (diffusion.py), per-atom element labels are embedded and passed directly as node features at every diffusion step; atom types are fixed inputs throughout both training and inference, and the model outputs only lattice and coordinate predictions.

\textbf{Element-Set Conditioning and Stoichiometry Inference.}
In the element-set variant, atom types are withheld. Instead, the ground-truth per-atom one-hot vectors are collapsed to a per-crystal multi-hot binary vector $\mathbf{e} \in {0,1}^{100}$ via a max-reduction, retaining only element identity and discarding counts. This vector is broadcast to every atom, linearly embedded, and concatenated with the sinusoidal time embedding before being projected to the working hidden dimension. The model additionally outputs per-atom type logits, trained with cross-entropy against the ground-truth labels. At inference, the element set is supplied directly as a pre-built multi-hot vector and kept fixed across all reverse-diffusion steps. Atom-type logits are produced at every decoder call but are only used from the final predictor evaluation at $t=1$, when the geometry is most accurate. Before the discrete assignment, logits corresponding to elements absent from the element set are masked and resolved to discrete assignments via argmax to produce the stoichiometry.

\subsection{Augmented Data} \label{appndx:aug_data}
We compare our peak-features representation against models trained on heavily augmented diffraction patterns. Following the augmentation strategy in \cite{riesel2024crystal}, we construct augmented versions of the Alex-MP-20-Poly and OQMD-Poly datasets designed to mimic experimentally observed peak broadening, intensity distortions, and background noise. 

\textbf{Peak-Shape Augmentation.}
Each augmented spectrum is re-simulated directly from the ground truth CIF using four different sets of Caglioti parameters $(U,V,W)$ (see SI Section~\ref{appndx:PXRD_rep} for details on the PXRD representation and Caglioti broadening). The four parameter configurations are $(0.05,-0.06,0.07)$, $(0.05,-0.01,0.01)$, $(0.0,0.0,0.1)$, and a random broadening variant in which $W$ is sampled uniformly from $[0.001, 0.1]$ while $U = V = 0$. Together, these configurations span a broad range of peak widths commonly encountered in powder diffraction experiments.

For all variants except $(0.05,-0.01,0.01)$, individual Bragg peak intensities are further perturbed by independent multiplicative Gamma-distributed factors. Specifically, each peak intensity $I_i$ is scaled as $I_i \leftarrow I_i g_i$, where $g_i \sim \mathrm{Gamma}(a,\mathrm{scale})$. The shape parameter $a$ and scale parameter were estimated via maximum-likelihood fitting to the distribution of per-peak observed-to-simulated intensity ratios extracted from the RRUFF evaluation subset. This augmentation is intended to mimic experimentally induced deviations in relative peak intensities, such as preferred orientation and other non-ideal scattering effects. Applying all four variants increases the size of each training set by a factor of four, with each augmented sample providing a spectrally distinct realization of the same underlying crystal structure.

\textbf{Dynamic Noise Augmentation.}
In addition to static peak-shape augmentation, dynamic background noise augmentation is applied online during training. At each forward pass, an additive background noise trace is independently generated for every spectrum in the batch and added to the input PXRD pattern.

The background noise for spectrum $i$ is sampled from a Gamma distribution parameterized by a shape parameter $a_i$ and scale parameter $s_i$. Rather than fixing these parameters globally, $(a_i, s_i)$ are sampled independently for each spectrum from an empirically calibrated bivariate log-normal distribution over $(\log a, \log s)$. The calibration statistics are estimated from the same non-evaluation RRUFF calibration subset used for the peak-intensity perturbation fitting. For each calibration spectrum, low-intensity regions of the diffraction pattern are used to approximate the experimental background, after which Gamma distributions are fitted to the resulting background intensities using maximum-likelihood estimation (MLE). A Gaussian distribution is then fit in log-parameter space to the collection of estimated Gamma parameters across all calibration spectra.

To avoid unrealistic artefacts, extreme parameter draws are rejected and resampled, and the generated noise traces are clipped to suppress rare stochastic spikes before being added to the spectrum. This procedure exposes the model to a diverse range of experimentally realistic background profiles during training.



\end{appendices}


\end{document}